\documentclass[namedreferences]{SolarPhysics}
\usepackage[optionalrh]{spr-sola-addons} 
\usepackage{graphicx}        
\usepackage{color}           
\usepackage{url}             
\usepackage{textcomp}

\newcommand{\eg}{{\it e.g.}}

\begin{document}

\begin{article}

\begin{opening}

\title{Observations of quasi-periodic solar X-ray emission as a result of MHD oscillations in a system of multiple flare loops}

\author{I.V.~\surname{Zimovets}\sep
               A.B.~\surname{Struminsky}     
       }
\runningauthor{I.V.~Zimovets, A.B.~Struminsky}
\runningtitle{Observations of quasi-periodic solar X-ray emission}

\institute{Space Research Institute, Russian Academy of Sciences, Profsoyuznaya str. 84/32, Moscow, 117997 Russia\\
email: \url{ivanzim@iki.rssi.ru}
}

\begin{abstract}
We investigate the solar flare of 20 October 2002. The flare was accompanied by quasi-periodic pulsations (QPP) of both thermal and nonthermal hard X-ray emissions (HXR) observed by RHESSI in the 3-50 keV energy range. Analysis of the HXR time profiles in different energy channels made with the Lomb periodogram indicates two statistically significant time periods of about 16 and 36 seconds. The 36-second QPP were observed only in the nonthermal HXR emission in the impulsive phase of the flare. The 16-second QPP were more pronounced in the thermal HXR emission and were observed both in the impulsive and in the decay phases of the flare. Imaging analysis of the flare region, the determined time periods of the QPP and the estimated physical parameters of magnetic loops in the flare region allow us to interpret the observations as follows. \textbf{1)} In the impulsive phase energy was released and electrons were accelerated by successive acts with the average time period of about 36 seconds in different parts of two spatially separated, but interacting loop systems of the flare region. \textbf{2)} The 36-second periodicity of energy release could be caused by the action of fast MHD oscillations in the loops connecting these flaring sites. \textbf{3)} During the first explosive acts of energy release the MHD oscillations (most probably the sausage mode) with time period of 16 seconds were excited in one system of the flare loops. \textbf{4)} These oscillations were maintained by the subsequent explosive acts of energy release in the impulsive phase and were completely damped in the decay phase of the flare. 
   
\end{abstract}
\end{opening}

\section{Introduction}
     \label{S-Introduction}

\indent
Electrons, accelerated during solar flares, generate nonthermal HXR and microwave emission in the solar atmosphere as a result of their Coulomb collisions with ambient plasma (bremsstrahlung) and interaction with magnetic field (synchrotron radiation), respectively. Often the light curves of nonthermal HXR and microwave emissions in large solar flares are composed of multiple pulses with different durations from less than one second up to several minutes (\eg,~ \opencite{dennis88};~\opencite{aschwanden09}, and references therein). This indicates many episodes of acceleration of electrons, possibly in different locations within flare regions, since the majority of these flares are accompanied by formation of arcades of many flaring loops and by the HXR emission generated in different flare loops (\eg,~ \opencite{vorpahl76};~\opencite{dejager79};~\opencite{grigis05}). It has been suggested that flares consist of a number of ``Elementary Flare Bursts'' (EFB), which can be a result of collisions between different current-carrying loops \cite{dejager79,sakai96}. 

\indent
More rarely, the light curves of nonthermal HXR and microwave emission show apparent quasi-periodic pulsations (QPP; see, \eg,~ \opencite{dennis88}; ~\opencite{nakariakov05};~ \opencite{aschwanden09} for a review and references therein). This may indicate the presence of some quasi-periodic processes (not fully understood at the moment) of acceleration of electrons. Using RHESSI imaging observations in the HXR range, it was shown that this quasi-periodic acceleration of electrons, at least in some flares, occurs in different flare loops, rather than in a single one \cite{zimovets09}. In this case the observed quasi-periodicity could be a coincidence owing to similar physical conditions in successively bursting flare loops. This returns us to the idea of multiple EFB due to coalescence of current-carrying loops or due to the presence of multiple spatially separated ``null-points'' in flare regions, in which the stored magnetic energy can be released. But in principle, it is possible to implement a model proposed by \inlinecite{nakariakov06} to explain the QPP in the latter case by the successive quasi-periodic reconnection in different null-points due to fast magnetoacoustic oscillations in nearby non-flaring loops.    

\indent
On the other hand, it was theoretically suggested that QPP could be produced inside a single vibrating flare loop by the quasi-periodic modulation of the spectrum of nonthermal electrons through the betatron acceleration \cite{brown75} or by the quasi-periodic modulation of efficiency of the trapped nonthermal electrons to precipitate into the lower heights with larger plasma density and stronger magnetic field (in particular, the sausage mode is a good candidate for this) \cite{zaitsev82}. Indeed, imaging observations of the QPP in the microwave range made using the Nobeyama Radioheliograph with sufficient spatial resolution (but with no simultaneous QPP observed in the HXR range), which could be interpreted in terms of the global sausage mode of the flaring loop oscillations, were reported \cite{nakariakov03}. Imaging observations of QPP, but without sufficient spatial resolution together with an analysis of the time periods of QPP, allowed QPP to be interpreted in terms of the flare loop oscillations (\eg,~ \opencite{asai01};~ \opencite{inglis09}). Different modes of oscillations in coronal loops actually have been found with the TRACE and SOHO spacecraft in the range of the extreme ultraviolet radiation (see, \eg,~ \opencite{aschwanden09};~ \opencite{nakariakov05} for a review and references therein), making the oscillations of magnetic loops a natural way to explain the QPP of electromagnetic emission in flares. However, despite the large number of papers published on flare QPP, their nature is still not fully understood. Further imaging analysis of flares with QPP is required. 

\indent
In this work we investigate the solar flare of 20 October 2002. The light curves of its thermal and nonthermal HXR emission detected by RHESSI clearly indicate the presence of the QPP with two significant time periods of about 16 and 36 seconds. We will present time-spectral, energy-spectral and imaging analysis of the flare HXR emission and propose possible flare scenario to interpret the observed QPP.  
\section{Data and Instrumentation} 
      \label{S-data}      

\indent 
\indent
\textbf{1.}	The \textit{Reuven Ramaty High-Energy Solar Spectroscopic Imager} (RHESSI; \opencite{lin02}) is used to detect solar X-rays of about 3-17000 keV. The time resolution of RHESSI is about 4 seconds, which equals the satellite's period of rotation. RHESSI can detect X-rays with energy resolution of about 1 keV \cite{smith02} and with spatial resolution of about $2.3^{\prime\prime}$ in the 3-100 keV range \cite{hurford02}.  

\indent
\textbf{2.}	Three-second data sets of X-ray emission detected by the \textit{X-Ray Sensor} (XRS) onboard GOES-10 satellite in two energy channels (1-8 and 0.5-4 \textrm{\r{A}} or about 1.55-12.4 and 3.1-24.8 keV) are used. Unfortunately, the Solar X-ray Imager onboard the GOES-12 satellite \cite{hill05} did not produce images of the Sun during the flare.

\indent
\textbf{3.}	One-second ground-based radio data (flux) are obtained from two observatories of the Radio Solar Telescope Network (RSTN) - the Sagamore Hill Radio Observatory (USA) and the San Vito Solar Observatory (Italy) - at standard frequencies 245, 410, 610, 1414, 2695, 4995, 8800 and 15400 MHz. Unfortunately, these observatories do not make imaging observations of the Sun. The \textit{Nobeyama Radioheliograph}, who can make images of the Sun in the microwave range, was in night time during the flare.

\indent
\textbf{4.}	Full-disk images made in Fe XII lines (about 195 \textrm{\r{A}}; formation temperature is about 1.6 MK) by the \textit{Extreme-ultraviolet Imaging Telescope} onboard the \textit{Solar and Heliospheric Observatory} spacecraft (EIT/SOHO; \opencite{delaboudiniere95}) with a spatial resolution of $2.62^{\prime\prime}$ and cadence of about 12 minutes are used to determine configuration of magnetic loops in the flare region. Unfortunately, the \textit{Transition Region and Coronal Explorer} (TRACE; \opencite{handy99}), which has better spatial resolution and fewer cadence, observed different active region during this flare.

\indent
\textbf{5.}	Level 1.8 full-disk photospheric line-of-sight magnetograms made by the \textit{Michelson Doppler Imager} onboard SOHO (MDI/SOHO; \opencite{scherrer95}) with a spatial resolution of about $2^{\prime\prime}$ are used to determine configuration of magnetic field and the magnetic inversion line in the flare region and to roughly estimate magnetic field in flare loops.        
\section{Observations} 
      \label{S-observations}      
      
\subsection{Time and spectral analysis of the flare hard X-ray emission} 
  \label{S-time}     
      
\begin{figure}    
   \centerline{\includegraphics[width=1.\textwidth, bb=0 10 590 270, clip=]{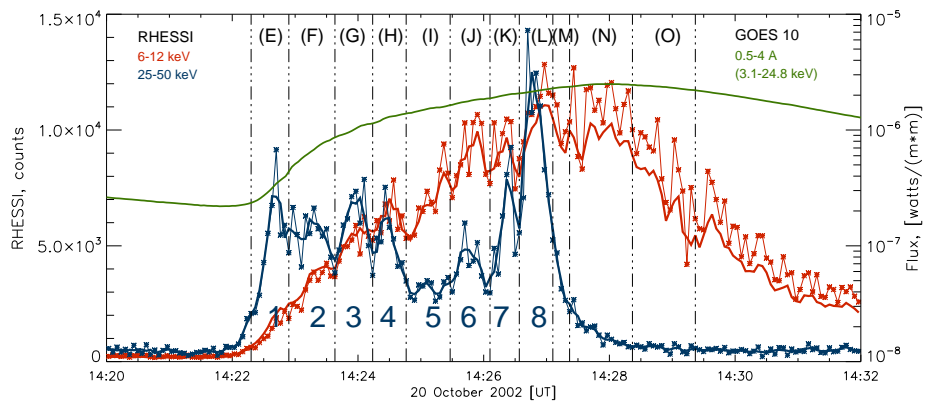}
              }
              \caption{Time profiles of the full-disk X-ray emission, observed by RHESSI's detectors 3, 4, 8 and by XRS/GOES-10 during the investigated solar flare. The time resolution of the RHESSI data is 4 seconds (thin blue and red lines with asterisk; corresponding thick solid lines are the RHESSI data smoothed over 12 seconds; left vertical axis in the linear scale). The time resolution of the GOES flux data is 1 second (green line; right vertical axis in the logarithmic scale). Bold blue numerals indicate each QPP with time period of about 36 seconds. The vertical dashed-dotted lines and the letters in parentheses indicate time intervals, in which the RHESSI images (presented in Figure~\ref{fig3} and denoted by the same, but small letters) are made. The impulsive phase is between (E) and (M), and the decay phase is after (M).
                      }
   \label{fig1}
\end{figure}

\indent
The solar flare of 20 October 2002 (M1.8 according to the GOES classification) with heliographic coordinates S19\textdegree{},~W23\textdegree{} started at about 14:21 UT in X-ray radiation in the NOAA Active Region 10163. The light curves of HXR emission detected by RHESSI in the 3-50 keV range indicate clearly the presence of QPP (Figure~\ref{fig1}). The light curve of X-ray radiation detected by GOES is much smoother and does not reveal pulsations. Possibly, this is due to RHESSI's detectors having much higher count statistics of X-rays than GOES's detectors: RHESSI has nine cryogenically cooled germanium detectors, while the GOES satellites have two ion chambers. Unfortunately, the used solar radio telescopes could not detect the above-background flux of microwave emission at frequencies above 8800 MHz, which could allow us to study the behavior of the mildly-relativistic electrons in this flare. Thus, further we will concentrate only on the analysis of HXR emission, detected by RHESSI.   

\indent
Using the OSPEX package within the SolarSoftWare we made spectral analysis of the HXR emission (from full-disk and from selected areas of the flare region) detected by RHESSI in different time intervals during the flare. The implemented best-fit-technique clearly indicates that the energy spectrum of the HXR emission has a thermal component and a nonthermal component, which is fitted well by the double power-law function (the break energy is in the 15-25 keV range; the power-law spectral indexes below and above break are between -9 and -4 during the flare; this soft spectrum of HXR emission partly explains the absence of detectable microwave emission). The nonthermal component dominates the thermal component in the spectrum of X-rays above energies of about 10-15 keV. Further, according to the implemented spectral analysis, we will subdivide the HXR emission into thermal (below 15 keV) and nonthermal (above 15 keV) components. For the sake of clarity and compactness of the paper, we will present an analysis of only light curves in the 5(6)-10(12) (thermal component) and 25-50 (nonthermal component) keV energy ranges.

\indent
Looking at the light curves of HXR emission, we can subdivide the flare into two fundamentally different phases: the impulsive phase and the decay phase. The impulsive phase consists of eight quasi-periodic pulses of nonthermal emission [episodes (E)-(L) on Figure~\ref{fig1}] with a duration of about 30-40 seconds, indicating several acts of energy release and acceleration of electrons. These pulses are clearly composed of shorter quasi-periodic subpulses with a duration of about 15-20 seconds, which coincide in time (within the measurement accuracy of RHESSI of about 4 seconds) with pulses of thermal emission. The decay phase, which started after the maximum of the thermal HXR emission [after interval (M) on Figure~\ref{fig1}], consists mainly of pulses of thermal emission with a duration of 15-20 seconds, while the decaying flux of nonthermal HXR emission mainly below 25 keV is also still presented (this is not shown on Figure~\ref{fig1} for clarity). Possibly, this indicates that there were no new significant acts of energy release in the decay phase of the flare, while nonthermal electrons, accelerated mainly in the impulsive phase, were still trapped in the ``post''-flare loops. This scenario will be confirmed by the imaging analysis of the flare region (see Section~\ref{S-images}).         

\indent
The presence of oscillations in the light curves of the observed HXR emission is confirmed by their spectral analysis. Firstly, to remove low frequency spectral components, we subtract the running averaged signals smoothed over 20 or 60 seconds $F_{20,60}(t)$ from the 4-second RHESSI data $F_{4}(t)$. Further, the normalized Lomb periodograms \cite{lomb76,scargle82} are calculated for the normalized signals $dF(t)/F_{20,60}(t)=\left[F_{4}(t)-F_{20,60}(t)\right]/F_{20,60}(t)$ in different time intervals:\\

\begin{figure}    
   \centerline{\includegraphics[width=1.\textwidth, bb=2 2 565 430, clip=]{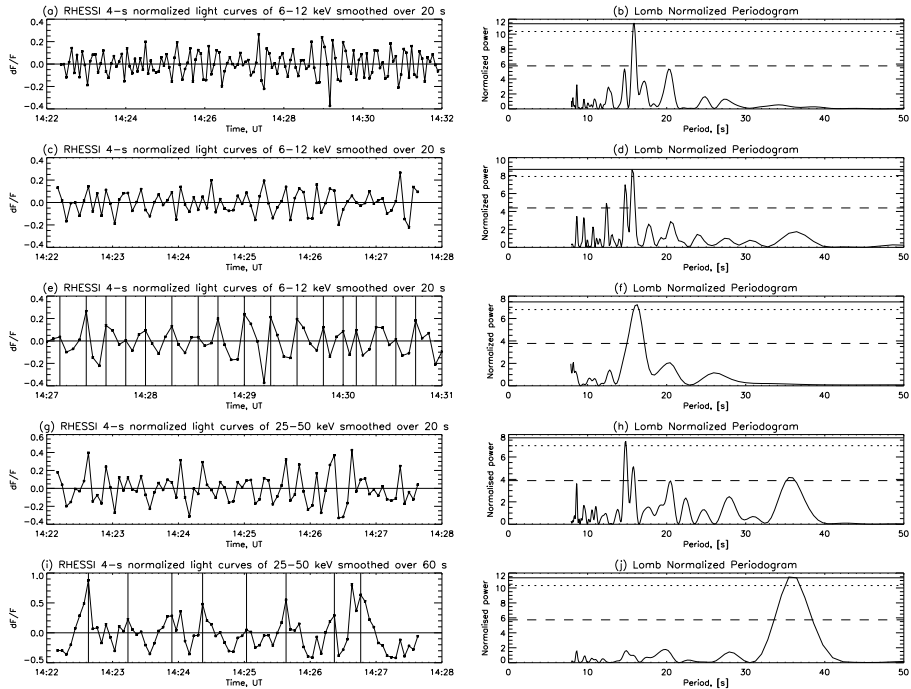}
              }
              \caption{Panels (a), (c), (e), (g) and (i) show normalized light curves of HXR emission detected by RHESSI in the 6-12 (thermal) and 25-50 (nonthermal) keV energy ranges in different time intervals of the 20 October 2002 solar flare: $dF(t)/F_{20,60}(t)=\left[F_{4}(t)-F_{20,60}(t)\right]/F_{20,60}(t)$. Panels (b), (d), (f), (h) and (j) show the normalized Lomb periodogram calculated for $dF(t)/F_{20,60}(t)$ on the Panels (a), (c), (e), (g), (i), respectively. The 50\%, 90\% and 99\% confidence levels are plotted by horizontal dashed, dotted and solid lines, respectively. Time profiles of thermal HXR emission, in particular in the 6-12 keV range, reveal oscillations with a main period of about 16 seconds both in the impulsive and in the decay phases of the flare. This signal is more periodic in the decay phase. Time profiles of nonthermal HXR emission, in particular in the 25-50 keV range, reveal two oscillation periods of 14-17 and 36-37 seconds.     
                      }
   \label{fig2}
\end{figure}

\indent
1) during the whole flare [Figure~\ref{fig2}(a)], in the impulsive [Figure~\ref{fig2}(c)] and in the decay phases [Figure~\ref{fig2}(e)] separately for thermal emission;\\

\indent
2) only in the impulsive phase for nonthermal emission (when there was statistically significant flux of X-rays above the RHESSI background level; [Figure~\ref{fig2}(g) and ~\ref{fig2}(i)].\\

\indent
The power spectra of the normalized light curves of thermal emission reveal a statistically significant \cite{horne86} peak of about 16 seconds both in the impulsive and in the decay phases, but more pronounced in the decay phase [see Figure~ \ref{fig2}(b), \ref{fig2}(d) and \ref{fig2}(f)]. The power spectra of the normalized light curves of nonthermal emission with averaging over 20 and 60 seconds reveal two statistically significant peaks around 14-17 seconds and about 36-37 seconds, respectively. Averaging over 60 seconds has strengthened the statistical significance of the 36-second peak in the spectrum, while averaging over 20 seconds has strengthened the significance of the 16-second peak. The splitting of the peak near 16 seconds to two more narrow peaks [Figure~\ref{fig2}(h)] may be related to the short duration of the analyzed data sets.    
        
\subsection{Imaging analysis of the flare region} 
  \label{S-images}   

\indent
Figure~\ref{fig3} shows images of the flare region made in different wavelength bands and in different time intervals during the flare. The EIT/SOHO images made in the impulsive phase and just after the decay phase jointly with an analysis of the MDI/SOHO photospheric magnetogram of the line-of-sight magnetic field [Figure~\ref{fig3}(a--d)] clearly indicate two closely spaced systems of flare loops: in the south-east (SE) and in the north-west (NW). Unfortunately, the EIT/SOHO cadence of about 12 minutes is not enough to investigate the dynamics of this flare in the EUV radiation.

\indent
To examine the spatial evolution of thermal and nonthermal X-ray sources during the impulsive and decay phases we reconstruct RHESSI images using the CLEAN and Pixon algorithms \cite{hurford02}. Detectors 3-8 are used. Several series of images integrated over 8, 12, 16, 20, 32, 48, 60 seconds are obtained in different energy ranges and for different time intervals. We mainly work with the 5(6)-10(12) keV (thermal X-ray sources) and 25-50 keV (nonthermal X-ray sources) RHESSI images.

\indent
Figure~\ref{fig3}(e--o) shows the morphology and dynamics of the flare region in X-rays. It is seen that early in the impulsive phase (up to about 14:25 UT) there were two clearly separated systems of flare loops (in the South-East and in the North-West), whose positions coincided well with the loop systems SE and NW, visible in the EIT images. These loop systems were not independent. We calculate time profiles of thermal HXR flux at 6-12 keV both from the SE and from the NW systems in the 14:22:48-14:25:20 UT interval (when the two flare systems were clearly separated), which we split into subintervals of 8 seconds to make time series of X-ray images. To calculate the HXR flux from each system we employ two circular integration regions with radius of 15 arcsecs for each system on each image. It turned out that fluxes of thermal HXR emission, calculated from two separate flare systems, were strongly correlated (Figure~\ref{fig4}).

\indent
The space between the two flare loop systems was filled with thermal HXR radiation after about 14:25 UT (before start of the decay phase), indicating the presence of magnetic loops here, which could connect the SE and the NW systems. The SE system and the space between two systems have become practically invisible in thermal HXR against the background of the bright NW source in the decay phase, possibly due to much stronger emission from the NW system and the limited dynamic range of the RHESSI observations. Only the NW system was clearly observed in thermal HXR in the decay phase. Unfortunately, it is not possible to make spatially resolved analysis of the flare oscillations in thermal HXR emission due to the low spatial resolution of RHESSI.  

\indent
Morphology of nonthermal HXR sources, located mainly in the footpoints of different flare loops, also confirms the presence of two flare loop systems before about 14:25 UT. We emphasize that nonthermal HXR were emitted from different positions in both flare systems during this time interval - the sources of nonthermal HXR emission were not stationary! The footpoint sources of nonthermal HXR emission almost completely disappeared in the SE system after about 14:25 UT [Figure~\ref{fig3}(j) and \ref{fig3}(k)], while two footpoint nonthermal HXR sources in the NW system became stationary. Nonthermal HXR radiation filled the whole NW loop system, not only its footpoints, in the decay phase [Figure~\ref{fig3}(m), ~\ref{fig3}(n) and ~\ref{fig3}(o)]. Possibly, this was caused by the trapped nonthermal electrons in these overdense loops. Unfortunately, it is not possible to investigate dynamics of nonthermal HXR emission more precisely due to low RHESSI count statistics in this flare (we have to use integration at least over 32 seconds to generate 25-50 keV images of high quality).

\section{Discussion} 
  \label{S-discussion}   

\begin{figure}    
   \centerline{\includegraphics[width=1.\textwidth, clip=]{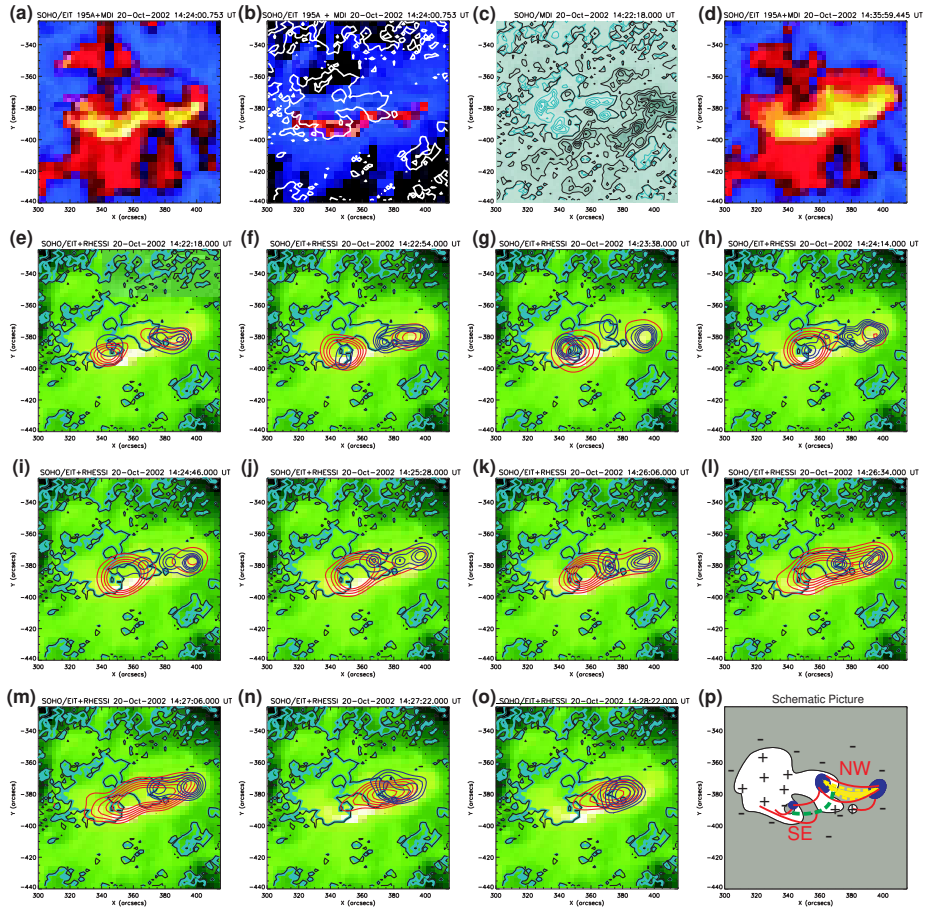}
              }
              \caption{Images of the 2002 October 20 solar flare made in different spectral ranges of electromagnetic emission (projection on the photosphere; the start time of each image is shown above). (a, d) EIT/SOHO images at 195 \textrm{\r{A}} made in the impulsive phase and just after the decay phase of the flare, respectively. (b) EIT running-difference 195 \textrm{\r{A}} image. White lines indicate the zero-gauss isocontours (\eg,~ the magnetic inversion lines) of the MDI photospheric magnetogram made in the impulsive phase. (c) MDI magnetogram of the flare region overlaid by the isocontours at levels $\pm$(10, 310, 610, 910, 1210, 1510) gauss (turquoise and black lines). (e--o) The green-color images are the same EIT 195 \textrm{\r{A}} image as shown in (d), but using another color palette for clarity. Turquoise and black lines are the isocontours of MDI magnetogram at levels $\pm$10 gauss indicating the magnetic inversion lines. Red and blue lines are the RHESSI 5-10 keV and 25-50 keV contours (30\%, 40\%,$\ldots$, 90\%) indicating thermal and nonthermal HXR sources, respectively. The RHESSI images, shown in panels (e--o), are made for the appropriate time intervals marked by the same letters in Figure~\ref{fig1}. (p) Sketch of the flare region. The photospheric line-of-sight magnetic field of positive/negative polarities is shown by white/grey colors ($\pm$ signs). Red lines, ``SE'' and ``NW'' designations indicate two systems of flare loops, which can interact through the oscillating green dashed line (the 36-second periodicity). Yellow solid and dashed lines indicate the oscillating flare loops in the NW system (the 16-second periodicity). Blue ellipses indicate the footpoint sources of nonthermal HXR emission.        
                      }
   \label{fig3}
\end{figure}

\begin{figure}    
   \centerline{\includegraphics[width=1.\textwidth, bb=2 2 375 190, clip=]{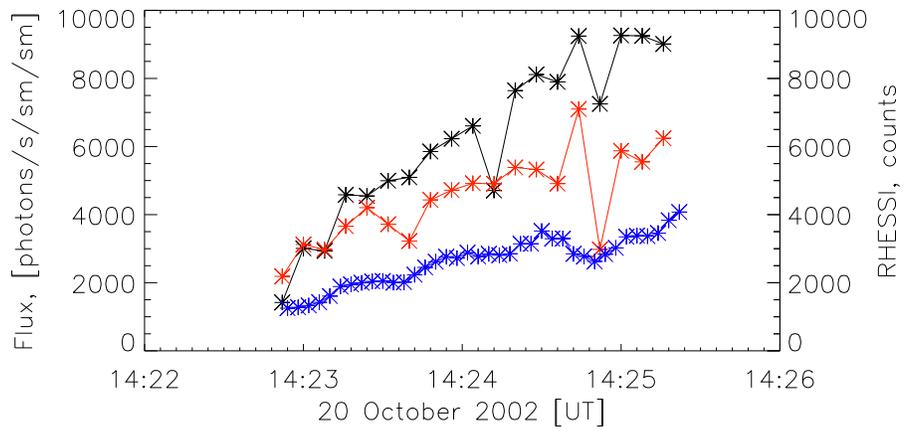}
              }
              \caption{Time profiles of the RHESSI HXR flux calculated from the SE (multiplied by 1.1) and from the NW (multiplied by 1.5) flare systems [see Figure~\ref{fig3}] using the Clean images at 6-12 keV reconstructed with the time cadence of 8 seconds (black and red lines with asterisk, respectively; left vertical axis). Time profiles of the full-disk X-ray emission, observed by RHESSI in the 6-12 keV range with the time resolution of 4 seconds and running-averaged over 8 seconds (blue line with asterisk; right vertical axis).
                      }
   \label{fig4}
\end{figure}

Based on the analyzed observational data, we can propose a rough scenario of the investigated flare [see Figure~\ref{fig3}(p)].

\indent 
\textbf{1.} \textbf{In the impulsive phase of the flare energy was released and electrons were accelerated by successive quasi-periodic acts with an average time period of about 36 seconds in different parts of two spatially separated, but interacting loop-systems of the flare region.} Possibly, these loop-systems interact through the connecting magnetic loop near both ends of which they are located. We have established two observational evidences that these flare systems are linked: 1) the fluxes of thermal HXR emission from these systems are strongly correlated; 2) the formation of the loop-type structure in the space between these flaring systems is observed in thermal HXR emission. The question appears: is it possible to interpret the observed 36-second quasi-periodicity ($P_{36}$) of energy release in two different flaring systems by the MHD oscillations in the connecting loop? Indeed, it was concluded from the imaging observations of another flare event that oscillations of even nonflaring transequatorial loop (the kink modes) can be responsible for the QPP of HXR emission observed simultaneously from two different active regions located at both ends of this loop \cite{foullon05}. Unfortunately, it is not possible to estimate length of the connecting loop ($L_{c}$) accurately in our case, due to insufficient spatial resolution of observations, nevertheless we can say with certainty that $20\leq{L_{c}}\leq40$ Mm. Thus, we can estimate the required phase speed of a fundamental standing mode as $V_{ph}=2L_{c}/P_{36}$ or $1100\leq{V_{ph}}\leq{2200}$ km/s. These phase speeds are consistent with the speeds of fast magnetoacoustic kink modes that have been directly observed in coronal loops in the EUV emission (\opencite{nakariakov05}; \opencite{aschwanden09}; and references therein). Implementing the imaging spectroscopy technique to the RHESSI HXR data we could estimate the emission measure of different parts of the flare region. Hence, using the observed volumes of these parts, we could roughly estimate averaged electron plasma density ($n_{e}$) in them. Our estimations give ${10^{10}}\leq{n_{e}}\leq{10^{11}}$ cm$^{-3}$ for each flaring loop-system and also for the connecting loop-type structure. Thus, the Alfv\'{e}n speed in the connecting loop-type structure is $7B\leq{V_{A}}\leq{22B}$ ~km/s, where $B$ ~is magnetic field in gauss inside the loop-type structure. Consequently, if $B\approx{100}$ ~G, which is a reasonable value, we have $700\leq{V_{A}}\leq{2200}$ ~km/s. These values of the Alfv\'{e}n speed are very close to the estimated $V_{ph}$! Thus, we conclude that the observed 36-second periodicity can actually be caused by the fast MHD oscillations in the loop-type structure, which connects two separate flare systems. The physical mechanism, proposed by \inlinecite{nakariakov06}, is a good candidate to explain quasi-periodic spatial fragmentation of energy release in the impulsive phase of this flare (see Section~\ref{S-Introduction}). Multiple null-points could actually exist in this flare region, because of its complex topology: several inclusions of magnetic field of opposite polarity are seen on the magnetograms.

\indent
\textbf{2.}	\textbf{During the first 36-second explosive episodes of energy release, the MHD oscillations, probably global sausage modes} (see below)\textbf{, with time period of about 16 seconds were excited in the loops of the NW flare system.}

\indent
\textbf{3.}	\textbf{These oscillations were maintained by the subsequent explosive acts of energy release in the impulsive phase and were damped in the decay phase of the flare.} We conclude that the 16-second oscillations were excited only in the NW system, but not in the SE flare system and not in the connecting loop-type structure at least in the decay phase, because the NW system was more stationary and retained its loop-like configuration in the decay phase, when the QPP of thermal HXR emission were still observed, but the SE flare system and the connecting loop-type structure were no longer visible in thermal HXR emission on the RHESSI images. Estimated length and electron plasma density of the loops in the NW flare system are $25\leq{L_{NW}}\leq{35}$ Mm and ${10^{10}}\leq{n_{e}}\leq{10^{11}}$~cm$^{-3}$, respectively. These physical parameters highly coincide with the same parameters, found in \inlinecite{nakariakov03}, where the 16-second quasi-periodic pulsations of microwave emission, observed with the Nobeyama Radioheliograph, were successfully interpreted in terms of the global sausage mode of the oscillating flare loop! Thus, by analogy, we can also interpret the observed 16-second pulsations of HXR emission as the global sausage mode, excited in the loops of the NW flare system. Moreover, the sausage mode must be accompanied by perturbations of plasma density and emission measure. Hence, this mode of oscillations would quasi-periodically modulate flux of thermal HXR emission. This is what we observe both in the impulsive and in the decay phase of the flare. The decay phase was not accompanied by significant energy release, therefore we clearly observe the damping of oscillations. Contrary to that, in the impulsive phase there were multiple explosive acts of energy release, which might intermittently re-excite oscillations.

\indent
It is interesting to note that the 16-second periodicity of nonthermal emission has already been observed in other solar flares \cite{parks69,nakariakov03,inglis08}. Perhaps, this is a typical period of the global sausage oscillations in flare regions \cite{inglis08}. The 36-second periodicity of HXR emission was also found in many solar flares \cite{lipa78}. 

\indent
Despite the above arguments we are aware that the proposed flare scenario, based on the MHD oscillations of magnetic loops in flare region, may not be true. Oscillations of the flaring loops itself were not detected directly using imaging observations, though this may be due to the observational limitations.

\begin{acks}
We are grateful to the spacecraft teams and consortia (RHESSI, \\ SOHO, and GOES) and ground-based observatories (the Sagamore Hill Radio Observatory and the San Vito Solar Observatory), whose data were used in this work. We thank Jenny Harris for help in English editing. IVZ is grateful to the organizers of the Baikal Young Scientists' International School for Fundamental Physics (2009), where this work was presented. This work was partially supported by the Russian Foundation for Basic Research, grants No. 07-02-00319, 09-02-16032.

\indent
This article is dedicated to IVZ's grandfather F.I.~Krylov, veteran of the WWII, who died during its writing.

\end{acks}
\clearpage{}

\bibliographystyle{spr-mp-sola-cnd} 
\bibliography{oscillations}  
\IfFileExists{\jobname.bbl}{} {\typeout{}
\typeout{****************************************************}
\typeout{****************************************************}
\typeout{** Please run "bibtex \jobname" to obtain} \typeout{**
the bibliography and then re-run LaTeX} \typeout{** twice to fix
the references !}
\typeout{****************************************************}
\typeout{****************************************************}
\typeout{}}

\newpage

\newpage

\newpage

\newpage


\end{article} 
\end{document}